\documentclass[aps,pr,twocolumn,showpacs,superscriptaddress,groupedaddress,nofootinbib]{revtex4}  

\usepackage{longtable}
\usepackage{graphicx}  
\usepackage{ulem}   
\usepackage{comment} 
\usepackage{datetime}
\usepackage{dcolumn}

\usepackage{amsxtra}
\usepackage{euscript} 
\usepackage{bm}        
\usepackage{tabularx}
\usepackage{float}
\restylefloat{table}

\usepackage[cmtip,arrow]{xy}
\usepackage{pb-diagram,pb-xy}

\usepackage{amsmath}
\usepackage{amssymb}
\usepackage{amsthm}
\usepackage[pdftex]{color}
\usepackage[pdftex,colorlinks,citecolor=blue,linkcolor=blue,urlcolor=blue]{hyperref} 
\usepackage{graphicx}
\usepackage{dcolumn} 
\usepackage{bm} 

\usepackage{longtable}

\usepackage{ulem}   
\usepackage{comment} 
\usepackage{datetime}

\usepackage{tabularx}
\usepackage{float}
\restylefloat{table}

\newcommand{\gsim}{\;\rlap{\lower 3.5 pt \hbox{$\mathchar \sim$}} \raise 1pt
\hbox {$>$}\;}
\newcommand{\lsim}{\;\rlap{\lower 3.5 pt \hbox{$\mathchar \sim$}} \raise 1pt
\hbox {$<$}\;}

\begin{document}
\title{\boldmath
An intriguing connection between Pisarski's fixed point and $(2+3)$-spin glasses 
\unboldmath}
\author{Vincent Lahoche}
\email[]{vincent.lahoche@cea.fr}
\affiliation{Université Paris Saclay, \textsc{Cea}, \textsc{List}, Gif-sur-Yvette, F-91191, France}
\author{Dine Ousmane Samary}
\email[]{dine.ousmanesamary@cea.fr}
\affiliation{Université Paris Saclay, \textsc{Cea}, \textsc{List}, Gif-sur-Yvette, F-91191, France}
\affiliation{Faculté des Sciences et Techniques (ICMPA-UNESCO Chair)\protect\\
Université d'Abomey-Calavi, 072 BP 50, Bénin}

\begin{abstract}
This paper aims to establish a connection between Pisarski's fixed point and a $(2+3)$-spin-glass model with sextic confinement potential. This is made possible by the unconventional power-counting induced by the effective kinetics provided by the disorder coupling in the large $N$-limit. Because of the absence of epsilon expansion, our approach is more attractive than the previous one. It may be relevant to the signal detection issue in nearly continuous spectra.


\end{abstract}
\maketitle

\begin{figure}
\begin{center}
\includegraphics[scale=0.6]{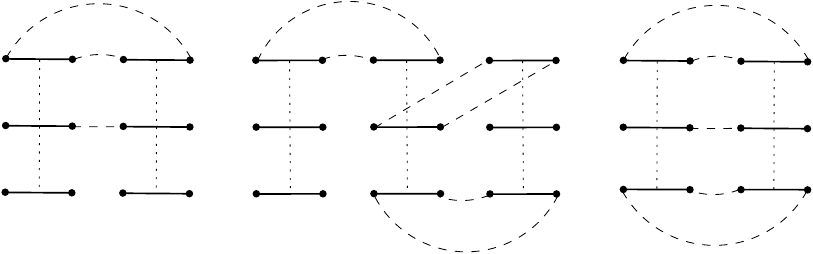}
\end{center}
\caption{Typically relevant two and four loops diagrams for the critical sextic theory, for effective coupling (diagrams one and two on left) and anomalous dimension (on right). Solid edges materialize scalar products $\vec{\phi}\cdot \vec{\phi}$, and vertices are sets of 3 solid edges linked with a dotted edge. Propagators are the dashed edges. A face is an alternate-closed sequence of dashed and solid edges.}\label{figdiagBMB}
\end{figure}

\paragraph{Introduction.} 
The perturbative Pisarski's fixed point, first discovered forty years ago, is an intriguing phenomenon in statistical field theory, particularly in the study of critical phenomena. In this phenomenon, the $\beta$-functions of the $\epsilon$-expansion of sextic $O(N)$ models near $3D$ have a non-trivial large $N$ limit. In this limit, for $D=3$, there exists an attractive fixed points line (FP) for the critical theory, starting from the Gaussian fixed point and ending at some special marginal (asymptotically safe) FP \cite{Pisarski,OSBORN}. To be more precise, Pisarski's phenomenon can be described shortly by the following argument. Consider a critical sextic $O(N)$ model, with classical action in dimension $D=3-\epsilon$:
\begin{align}
\nonumber S_{\text{cl}}[\vec{\phi}\,]\,=\,&\frac{1}{2}\int\, d^D \textbf{x}\,  \,\vec{\phi}(\textbf{x})\cdot (-\Delta) \vec{\phi}(\textbf{x})\\
&\qquad +\,\frac{g}{3N^2}\int\, d^D \textbf{x}\, (\vec{\phi}\,^2(\textbf{x}))^3\,,
\end{align}
where $\textbf{x}\in \mathbb{R}^D$, $\vec{\phi}:=\{\phi_1,\cdots,\phi_N\}$, and $\Delta$ is the standard Laplacian over $\mathbb{R}^D$. 
The perturbative expansion as usual is self-organized in a loop expansion but due to the special scaling of the sextic interaction, the loop expansion stops at four loops\footnote{Strictly speaking, this is at next to leading order, the loop corrections with $V=2,3$ are of order $N^{-3}$, while the bare coupling is of order $N^{-2}$.}. Indeed, a general Feynman amplitude $A(G)$ for some 1PI diagram $G$ scales as $A(G)\sim N^{-2V+F}$, where $V$ is the number of vertices involved in the corresponding diagram, and $F$ is the number of faces i.e the closed cycles of fields indices, sharing a factor $N$ (see Fig. \ref{figdiagBMB}). Furthermore, imposing the dimensional regularization \cite{ZinnJustinBook2}, the tadpoles diagrams must vanish $\int \, d^D \textbf{p} \, (\textbf{p}^2)^\alpha=0$.
Then, considering only the 1PI diagrams, a moment of reflection shows that the number of faces in the leading order (LO) graphs depends on the fact that $V$ is 
\begin{equation}
F_{LO} = \left\{
    \begin{array}{ll}
        \frac{3V}{2}-2 & \mbox{if } \, V=2n\\
        \frac{3(V-1)}{2} & \mbox{if}\, V=2n+1\label{Facescounting}
    \end{array}
\right..
\end{equation}
Then, we find that $A(G)\sim N^{-3}$ for $V=2,3$, and as $A(G)\sim \mathcal{O}(N^{-4})$ with $V>3$. The computation of the $\beta$-function leads to, taking into account similar quantum corrections for anomalous dimension:
\begin{equation}
N \beta(g)=-2N \epsilon g+12 g^2-\frac{\pi^2}{2}g^3 +\mathcal{O}(1/N)\,,
\end{equation}
which has two fixed-point solutions: $g_\pm=2 \left(6\pm\sqrt{36-\pi ^2 N \epsilon }\right)/{\pi^2}$. The negative sign branch corresponds to a large $N$ IR fixed point but collapses toward the Gaussian fixed point as $\epsilon=0$. In contrast, the positive sign branch is a nontrivial limit and corresponds to an interactive UV fixed point. By fixing the value of $\epsilon$, the fixed point does not make sense in the large $N$ limit, except if we assume that $N \epsilon= \mathcal{O}(1)$ i.e $\epsilon:=\alpha/N$, and the critical line seems to exist only for dimension $D\in [D_c(N),3]$, where $D_c(N):=3-\pi^2/36N$. Pisarski's computation is based on perturbation theory and two years after its publication, Bardeen, Moshe and Bander (BMB) showed that the UV Pisarski's FP lies in an unstable region of the phase space (where the one-loop effective potential is not bounded from below), but that the ultraviolet structure of the theory is governed by another fixed point, below the  Pisarski one \cite{bardeen1984spontaneous,David1,David2}. Since the last decade Pisarski's FP and the so-called \textit{BMB phenomenon} have been explored and generalized for different fields theories involving for instance fermions and supersymmetry \cite{Suzuki,Suzuki2} and a finite $N$ origin of BMB phenomenon has recently been investigated using nonperturbative renormalization \cite{Fleming,Yabunaka}. Note that both Pisarski's fixed point and BMB phenomenon are widely used in the literature. However, the connection we point out in this paper with $(2+3)$-spin glass is a novelty that requires special attention. Furthermore, to our knowledge, the first example in which the phenomenon occurs naturally in constructing the large $N$ limit, without requiring physically disputable methods like dimensional regularization is given in \cite{Schonfeld}.
\medskip

\paragraph{The model.} The first model we consider in this letter is a soft $2$-spin model, corresponding to the equilibrium probability distribution of the stochastic process:
\begin{equation}
\frac{d q_i}{dt}=-\sum_j W_{ij} q_j- \partial_i V(\vec{q}\,)+\eta_i(t)\,,
\end{equation}
where $\vec{q} \in \mathbb{R}^N$, $\vec{q}\,^2$ is the Euclidean square length, $W_{ij}$ are the entries of a $N\times N$ Wigner symmetric matrix of variance $\sigma^2$, and $\eta_i(t)$ is a Gaussian white noise $\langle \eta_i(t) \eta_j(t^\prime) \rangle = 2T \delta_{ij} \delta(t-t^\prime)$. As in \cite{van2010second,lahoche1}, the potential $V(\vec{q}\,)$  includes a disorder contribution and is designed to avoid arbitrary large spin configurations:
\begin{equation}
V(\vec{q}\,)= \sum_{n=1}^{M}\, \frac{a_n (\vec{q}\,^2)^n}{(2n)!N^{n-1}}\, +\sum_{i_1<\cdots < i_r} J_{i_1\cdots i_r} q_{i_1}\cdots q_{i_r}\,,
\end{equation} where  $M$ is the cutoff in the expansion series of this potential and  $J_{i_1\cdots i_r} $ is a random tensor with centered independent entries and correlation function :
\begin{equation}
\overline{J_{i_1\cdots i_r}J_{i_1^\prime \cdots i_r^\prime}}=\left(\frac{2\kappa r!}{(2r)! N^{r-1}}\right) \,\prod_{\ell=1}^r \delta_{i_\ell i_\ell^\prime}\,,\label{averageJ}
\end{equation}
where $\kappa$ is an arbitrary parameter.
This model can be studied analytically in the large $N$ limit. For $\kappa=0$, the model corresponds to the standard $2$-soft spin model. In this case, despite that the equilibrium dynamics solution does not exhibit a true spin glass phase and is nothing but a ferromagnet in disguise, the out-of-equilibrium problem is not so trivial and exhibits aging effects, for instance, \cite{Cugliandolo,Dominicis,lahoche1}. For $\kappa\neq 0$, the spherically constrained model can be solved as well using the replica technique \cite{Crisanti} at equilibrium. The equilibrium distribution $P_{\text{eq}}(\vec{q}\,)$ for given samples of $J$ and $W$ is $P_{\text{eq}}(\vec{q}\,)\propto \exp - H(\vec{q}\,)/T$, with $H(\vec{q}\,):= \frac{1}{2}\sum_{i,j} q_{i}W_{ij} q_j+V(\vec{q}\,^2)$. In the rest of this paper, we set $T=1$. First, let us focus on the case $\kappa=0$. For $N$ large enough, the spectrum for $W$ (assuming to be symmetric with i.i.d entires in the upper diagonal) becomes deterministic and goes toward the Wigner distribution \cite{Potters}\footnote{Note that the results of this article can be generalized to other distributions, such as the Marchenko–Pastur distribution for example. Similar results appear for any distribution whose tail follows the same power law as the Wigner distribution.} $\mu_W(\gamma)=\sqrt{4\sigma^2-\gamma^2}/2\pi \sigma^2$. We furthermore introduce the positive ‘‘generalized momenta'' \cite{lahoche3}: $p:=\gamma+2\sigma$, so that up the mass translation $\mu_1:= a_1-2\sigma$, the Hamiltonian $H(\vec{q}\,)\asymp H_{\infty}(\vec{\phi}\,)$, with:
\begin{equation}
H_{\infty}(\vec{\phi}\,):=\frac{1}{2} \sum_p \phi(p) (p+\mu_1) \phi(p) +\sum_{n=2}^{M}\, \frac{a_n (\vec{\phi}\,^2)^n}{(2n)!N^{n-1}}\, \,.\label{Hinfty}
\end{equation}
where $\phi(p)\asymp\sum_i q_{i}u_i^{(\gamma)}\vert_{\gamma=p-2\sigma}$, for some normalized eigenvector $u_i^{(\gamma)}$ of $W$, and $\vec{\phi}\,^2:=N \int d p \rho(p) \phi^2(p)$, where $\rho(\gamma+2\sigma)\equiv \mu_W(\gamma)$. Then, in the large $N$ limit, the equilibrium distribution looks like an ordinary field theory, with nearly continuous momenta $p$ and partition function:
\begin{equation}
Z_{\infty}[j]:=\int [d\phi] \, e^{-H_{\infty}(\vec{\phi}\,)+\sum_p \phi(p) j(p)}\,.\label{partition1}
\end{equation}

\begin{figure}
\begin{center}
\includegraphics[scale=0.7]{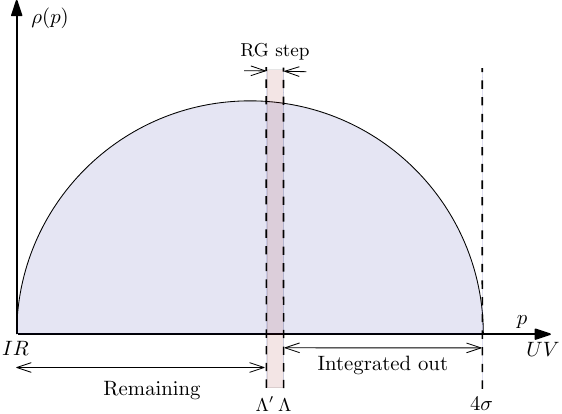}
\end{center}
\caption{A typical RG step on the Wigner spectrum.}\label{RGstep}
\end{figure}

\paragraph{Scaling and RG.} Following \cite{lahoche20241,lahoche3,Bradde}, an unconventional Wilsonian RG \cite{Wilson} can be constructed by partially integrating out degrees of freedom in the partition function \eqref{partition1}, from UV scales (large $p$) to IR scales (small $p$) -- see Fig. \ref{RGstep}. Note that there are four differences with ordinary RG: 1) The spectrum is not a power law like in ordinary QFTs. 2)  The spectrum is bounded. 3) The interactions are non-local in the usual sense \footnote{Such a kind of non-locality has been encountered in the literature \cite{Carrozza1}, and a suitable notion of locality as the RG can be defined in this context \cite{lahocheEVE}}. and 4) The sums in the Hamiltonian $H_{\infty}$ are without dimension. Accordingly with \cite{lahocheEVE,lahoche3}, a canonical notion of dimension can be defined for coupling regarding the behavior of the RG flow. Indeed, fixing some cut-off $\Lambda$ on the spectrum, $\mu_1$ scales as $\Lambda$ (i.e. like generalized momentum) under distribution dilatation. 
\medskip 

\begin{figure}
\begin{center}
\includegraphics[scale=1.2]{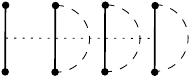}
\end{center}
\caption{LO contribution of order $a_4$ to the self energy.}\label{figureSigma}
\end{figure}

Now, let us consider the quantum correction to the self-energy of order $a_n$ (Fig. \ref{figureSigma}). The corresponding Feynman amplitude scale as $A(G)\sim (L_1(\Lambda))^{n-1}$, where $L_1(\Lambda)\sim \int_0^\Lambda \rho(p)/\Lambda $, and we define the \textit{canonical scaling} $I_n(\Lambda)$ of the coupling $a_n$ such that:
\begin{equation}
I_n(\Lambda) (L_1(0,\Lambda))^{n-1} \sim \Lambda\,,
\end{equation}
and the \textit{canonical dimension} as:
\begin{equation}
d_n:=\Lambda \frac{d}{d \Lambda}\, \ln\left(\Lambda (L_1(0,\Lambda))^{1-n} \right)\,.
\end{equation}
For the Wigner law, we get explicitly for the canonical dimension $d_n^{(0)}=(3-n)/2$ for $\Lambda$ small enough ($\forall \, \sigma$). This power counting matches with the standard power counting of a 3D field theory,  because for $p$ small enough, $\rho(p)\sim \sqrt{p}$. In addition with $\Lambda$ corrections arising because of the shape of the Wigner laws, there are also corrections  occurring by the $1/N$ expansion of the Wigner law \cite{Dhesi}, and $d_n$ looks like a series by fixing $\Lambda$:
\begin{equation}
d_n=d_n^{(0)}+\frac{1}{N}\, d_n^{(1)}+ \mathcal{O}(1/N^2)\,.
\end{equation}
We get in the deep IR, for $\sigma=0.5$\footnote{The integrals have to be suitably regularized because of the IR divergences.}, $d_n^{(1)}=\frac{791 \Lambda  (n-1)}{3645}+\mathcal{O}(\Lambda^{3/2})$ and $d_n^{(0)}=\frac{3-n}{2}+\frac{3}{20}(n-1)\Lambda+\mathcal{O}(\Lambda^{3/2})$. 
\medskip

\paragraph{Perturbative $\beta$-functions.} In \eqref{Hinfty} we set $M=3$ and we compute the LO $\beta$-function accordingly with the standard Pisarski's argument. Indeed, the power counting shows that quadratic and quartic couplings are relevant, and counter-terms could be defined in perturbation theory to fine-tune these couplings to zero, following the computation done in \cite{Pisarski}, and the relevant Feynman diagrams for 1PI 6 and 2-points functions for sextic couplings and anomalous dimension are again given by Fig. \ref{figdiagBMB}. To simplify the computation, we decided to use the standard Wetterich formalism\footnote{This choice allows easier contact with our work in preparation using the nonperturbative renormalization group. Other methods could also be used without any ambiguity.} \cite{Berges,lahoche20241} by adding a regulator $\Delta_\Lambda:=\frac{1}{2}\sum_p \phi(p) R_\Lambda(p) \phi(p)$ in the effective action, and in the deep IR $R_\Lambda(p)$ is equivalent to the Litim type regulator\footnote{This expression holds only in the deep IR, and has to be modified in the deep UV because of the boundary conditions for the effective average action, see \cite{lahoche20241}.} $R_\Lambda(p) \sim (\Lambda-p)\theta(\Lambda-p)$, where $\theta$ is the standard Heaviside function \cite{Litim}. All the diagrams involves powers of the $1$-face integral ($\mu_1\ll 1$):
\begin{equation}
I_2(\mu_1,\Lambda):=\int_0^2\, dp\, \frac{\rho(p)}{(p+\mu_1+R_\Lambda(p))^2}\,.
\end{equation}
The typical spacing between eigenvalues is of order $1/N$ i.e. of the same order as the effects we are aiming to compute. We can therefore use Wigner distribution in our computation, and we get (assuming again $\Lambda \ll 1$):
\begin{equation}
I_2(\mu_1,\Lambda)=\frac{16 \sqrt{2}}{3 \pi  \sqrt{\Lambda }}+\mathcal{O}(\Lambda^0,\sqrt{\mu_1})\,.
\end{equation}
In the rest, we use the standard mass scheme to compute RG equations. Because of the critical condition, the zero momenta $6$-points effective vertex function can be computed as:
\begin{equation}
N^3 \Gamma_\Lambda^{(6)}= N a_3-A(\Lambda) a_3^2+ B(\Lambda) a_3^3+\mathcal{O}(N^{-1})\,,\label{phi6}
\end{equation}
and furthermore, the wave function renormalization $Z$ is defined from the self energy $\Sigma(p)$ as the first derivative with respect to the external momenta $p$, for $p=0$: $Z:=1-\Sigma^\prime(0)=1-C(\Lambda) a_3^2N^{-2}+\mathcal{O}(N^{-3})$. From perturbation theory we get: $A=I_2/(10 \Lambda)$, $B=I_2^3/900$ and $C=0$. Note that the vanishing of $C$ is a consequence of the choice of the regulator. Then, equation \eqref{phi6} defines the effective coupling $u_3=a_3-(A a_3^2-B a_3^3)N^{-1}+\mathcal{O}(N^{-2})$. We then define the dimensionless coupling $u_3=:\bar{u}_3 I_3(\Lambda)$, and the $\beta$ function follows \cite{ZinnJustinBook2}: $\beta:= \Lambda \frac{d \bar{u}_3}{d\Lambda}$. To compute the RHS of this relation, we have to keep in mind that we focus on the deep IR regime $\Lambda \ll 1$. In particular, $I_3(\Lambda)= \pi^2/32+\mathcal{O}(\Lambda)$, computed with the Wigner law. We get:
\begin{align}
\nonumber\beta=&-\left(\frac{3}{10}+\frac{1}{N}\frac{1582 }{3645} \right) \Lambda \bar{u}_3 \\
&+ \frac{\sqrt{2}\pi}{40}\bar{u}_3^2 \left[1-\frac{8\bar{u}_3}{405}\right]\frac{\Lambda^{-3/2}}{N}+\mathcal{O}((\Lambda N)^{-2})\,.
\end{align}
 To understand the meaning of this $\beta$-function, we have to keep in mind that the typical spacing between eigenvalues is $\delta\sim N^{-1}$ (see \cite{Schlein,Potters}), and it is also the order of the minimum value for $\Lambda$ i.e. the last mode integrated out (see \cite{Lahochebreak}). Then, the construction of the IR limit has a non-trivial crossover with the large $N$ expansion, and we set $\Lambda = c N^{-\alpha}$ in the deep IR, assuming $c=\mathcal{O}(1)$ and $\alpha \leq 1$, the parameter $\alpha$ we introduced will allow us to characterize three different scaling regimes. Using \eqref{Facescounting}, we get that Feynman amplitudes scales as $A(G) \sim N^{\omega(G)}$ with:
\begin{equation}
\omega(G)=\frac{(3\alpha-2)V(G)-4\alpha+(1+\alpha) \theta_{V=2n+1}}{4}\,,
\end{equation}
where $\theta_{A}=1$ if $A$ is true and $0$ otherwise. Hence, as $\alpha \leq 2/3$, the contributions of higher order diagrams decrease in the large $N$ limit. Another interesting point is for $\alpha=2/5$, for which the dimensional and LO loop contributions have the same size (i.e. $\alpha=-3\alpha/2+1$). For $\alpha \leq 2/3$, we find a \textit{marginal fixed trajectory}:
\begin{equation}
\bar{u}_3^{(\pm)}=\frac{405}{16} \left(1\pm \sqrt{1-\frac{64 \sqrt{2} c^{5/2}}{135 \pi}N^{1-5\alpha/2}}\,\right)\,.
\end{equation}
Note that it is not a conventional fixed line, 
because of its scale dependency, but the $\beta$ function vanishes along it. Furthermore, note that there are no global fixed point because of the scale dependency of the $\beta$-function, as pointed out in \cite{lahoche3}. Finally marginal has here the usual sense in RG literature: the critical exponent vanishing for large $N$, for $2/5<\alpha < 2/3$.
\medskip

For $\alpha\geq 2/5=0.4$, the fixed point line is arbitrary close to $\bar{u}_3^*:=40.5$ as $c=\mathcal{O}(1)$, and $c \ll  (\pi\sqrt{2}/192)^{\frac{2}{5}}N^{\alpha-\frac{2}{5}}$. Finally, for $\alpha < \alpha_c(N)$, the fixed point solution break down, with:
\begin{equation}
\alpha_c(N):= \frac{2}{5}-\frac{K}{\ln N}\,,\quad K:=\ln \left(\frac{135 \pi}{64 \sqrt{2}} c^{5/2} \right)\,.
\end{equation}
Above this value, the dimensional contribution $-3 \Lambda \bar{u}_3/10$ dominates the flow and $\bar{u}_3 \sim \exp \left(-\frac{3}{10} \Lambda\right)$ (exponential regime). Fig. \ref{figureBMB} summarizes the different regions. Note that, as $\alpha>2/3$ (nonperturbative regime) higher order contribution cannot be discarded, and nonperturbative techniques are required to prove whether the fixed line exists or not.

\begin{widetext}
\begin{subequations}
\begin{figure}
\begin{center}
\includegraphics[scale=1.2]{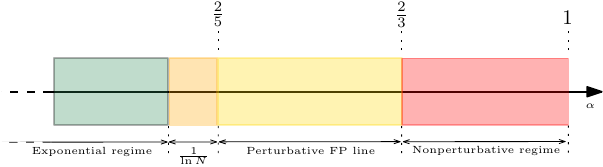}
\end{center}
\caption{Summary of the different IR regimes}\label{figureBMB}
\end{figure}
\end{subequations}
\end{widetext}

\begin{figure}
\begin{center}
\includegraphics[scale=0.9]{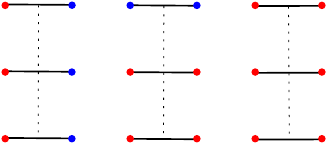}
\end{center}
\caption{The three relevant effective sextic interactions, that we denote respectively as $\lambda$, $\lambda^\prime$, and $u_3$ (as previously).}\label{figint}
\end{figure}

\begin{figure}
\begin{center}
\includegraphics[scale=0.6]{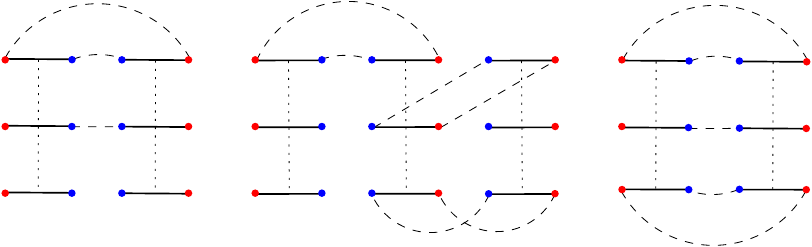}
\end{center}
\caption{Relevant contributions for $\beta_\lambda$ and $\beta_{\lambda^\prime}$ (respectively in the middle and the left) and for $\Sigma$ (on the right). Nodes marked with the same color have the same replica index.}\label{figdiagBMB2}
\end{figure}

\paragraph{(2+3)-spin glasses.} Let $\overline{Z_\infty^n}$ be the averaging of the replicated partition function \cite{Tarjus,Dominicis,Parisi}:
\begin{widetext}
\begin{eqnarray}
\overline{Z_\infty^n}=\int \prod_\alpha [d\phi_\alpha]\exp \left(-\sum_{\alpha=1}^n H_\infty(\vec{\phi}_\alpha)-W_\infty+\sum_{p,\alpha} \phi_\alpha(p) j_\alpha(p)\right)\,,\label{classicalaveraged}
\end{eqnarray} 
\end{widetext}
where Greek letters are for replica indices, and the averaging of the three body interaction reads:
\begin{equation}
W_\infty:=-\frac{\kappa}{6! N^{2}}\, \sum_{\alpha,\beta}\sum_{p}\, (\phi_\alpha(p) \phi_\beta(p))^{3}\,.
\end{equation}
Once again, we consider the critical regime, where quartic and quadratic local counter-terms are adjusted to cancel the corresponding asymptotically relevant observables. Note that the replica symmetry is explicitly broken in the expression above because of the source terms\footnote{Usually, replicas are introduced to compute the quenched averaging of $\ln Z$, taking the limit $n\to 0$ in $(Z^n-1)/n$. In this paper, we choose to break explicitly the replica symmetry, accordingly with reference \cite{Tarjus2}.\cite{Tarjus,Dominicis}.}. To deal with multi-replica vertices, we change our notation and adopt the convention that nodes marked with the same color on a vertex or on a graph have the same replica indices.  Fig. \ref{figint} shows the three interactions we can construct at the leading order, with three faces (all of them scale as $N^{-2}$). At the leading order, new couplings involving multi-replica vertices are perturbatively generated, as Fig. \ref{figdiagBMB2} shows, and we denote them as  $\lambda$ and $\lambda^\prime$. The $\beta$-functions can be computed from the same method as before, and with some effort, we get $Z=1$ (note that the bare notation is for dimensionless couplings):

\begin{align}
\nonumber\beta_{u_3}&=\beta\,+\, \frac{\sqrt{2}\pi}{40} \frac{\Lambda^{-3/2}}{N} \left(\frac{\bar{\lambda}^2}{2}-2\bar{\lambda}\bar{u}_3\right)\\
&-\frac{\sqrt{2}\pi}{2025}\frac{\Lambda^{-3/2}}{N}\left(\frac{\bar{\lambda}^3}{4}+3\bar{\lambda}^2\bar{u}_3-3\bar{\lambda}\bar{u}_3^2\right)+\mathcal{O}((\Lambda N)^{-2})\,,
\end{align}
\begin{equation}
\beta_\lambda=-\frac{3}{10}\Lambda\bar{\lambda}+\frac{\sqrt{2}\pi}{2025}\frac{\Lambda^{-3/2}}{N}\frac{\bar{\lambda}^3}{16}+\mathcal{O}((\Lambda N)^{-2})\,,\label{betalambda}
\end{equation}
\begin{equation}
\beta_{\lambda^\prime}=-\frac{3}{10}\Lambda\bar{\lambda}^\prime+\frac{\sqrt{2}\pi}{80} \frac{\Lambda^{-3/2}}{N} \bar{\lambda}^2+\mathcal{O}((\Lambda N)^{-2})\,.
\end{equation}
\medskip 

The equation \eqref{betalambda} can be easily solved, and in the windows of momenta $\alpha\in [\alpha(N),2/3]$, $\bar{\lambda}$ is almost constant for $N$ large enough $\bar{\lambda}\approx\bar{\lambda}(0)/0.55$, the initial condition being assumed in the IR regime, so that the equation \eqref{betalambda} makes sense.  Note that we have to keep in mind that the initial condition for $\bar{\lambda}$ assumes $\bar{\lambda}(0)<0$, but we will look at the state of the phase space in its entirety, including also the positive region. This region no longer has an interpretation in terms of a theory with disorder of type $2+3$, but could be accessible by nonperturbative methods beyond the scope of this article. We will therefore study it, anticipating more in-depth investigations.
\medskip

\begin{figure}
\begin{center}
\includegraphics[scale=0.5]{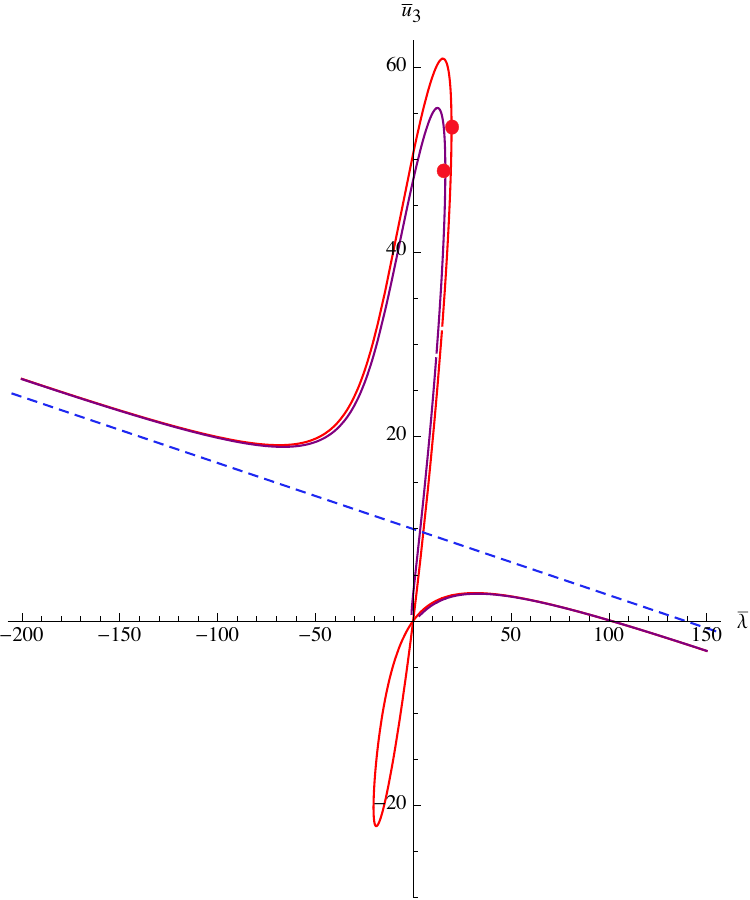}
\end{center}
\caption{Fixed lines in the space $(\bar{u}_3,\bar{\lambda})$, in the large $N$ limit, for $\alpha=1/2$ (red curve) and $\alpha=2/5$ (purple curve).}\label{FPLlambda}
\end{figure}

\begin{figure}
\begin{center}
\includegraphics[scale=1]{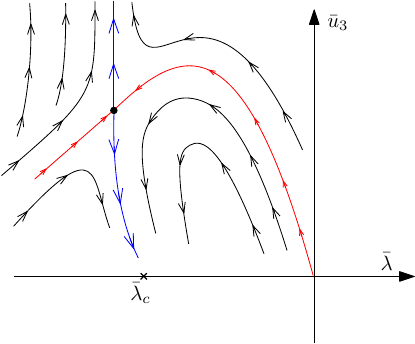}
\end{center}
\caption{Qualitative picture of the phase space around the Wilson-Fisher in the negative region for $\bar{\lambda}$. Before some $\bar{\lambda}_c(N)$, above the blue line, the flow reaches a regime where $\bar{\lambda}$ is almost frozen but $\bar{u}_3>0$.}\label{phasespace}
\end{figure}

Analyzing the zeros of $\beta_{u_3}$, we get three fixed lines in the large $N$ limit (but focusing on the interval $\alpha\in [\alpha(N),2/3]$), depending on the value of $\bar{\lambda}$ as Fig. \ref{FPLlambda} shows. In Fig. \ref{Expolines} we show the corresponding critical exponents and their dependency on $\bar{\lambda}$. The red line corresponds to the critical exponents along the unbounded fixed line. It is in the positive region, and the line is repulsive. The two last curves correspond to the critical exponents along the two pieces of the bounded fixed lines. The upper part (until the red dot on Fig. \ref{FPLlambda}) corresponding to the purple curve in Fig. \ref{Expolines} has a positive exponent and is global \textit{repulsive,} and the second part below the dot (blue curve) has a negative exponent and behaves as a worldwide \textit{attractor}. The repulsive line introduces a discontinuity in the phase space, reminiscent of a first-order phase transition, and the discontinuity point marked by the red dot identifies with a critical point. 
\medskip

Finally, analyzing the global zeros of the $\beta$ functions, we get a global fixed line, for:
\begin{equation}
\bar{\lambda}=-18 \sqrt{\sqrt{2}\frac{15}{\pi }} N^{\frac{1}{2} \left(1-\frac{5 \alpha }{2}\right)}\,.
\end{equation}
Each point of this line is a Wilson-Fisher-like fixed point, with one relevant direction and one irrelevant direction (see Fig. \ref{phasespace}). For $N$ large enough and $\alpha=2/3$, the critical exponent along the relevant direction is found numerically almost independent of $N$; we have for instance  $\theta_{\text{WF}}\approx 5.18$ for $N=10^5$ and the large $N$ limit value is numerically estimated as $\theta_{\text{WF}}^{\infty}\approx 5.62$. Note that the $\bar{u}_3$ coupling is nonperturbative. We find numerically $\bar{u}_3\sim 50$. For the irrelevant direction, the critical exponent behaves as $-3 N^{-\alpha}/5$, and goes to $0^-$ for $N$ large enough faster than the fixed point goes to $0$ (for $2/5<\alpha < 2/3$), meaning that the direction tends to begin marginal. Then the blue line looks like a separator line between a regime with vanishing $\bar{\lambda}$ (ordered phase) and a regime with non-vanishing $\bar{\lambda}$ (disordered phase), where replica symmetry breaking is enforced.

\begin{widetext}
\begin{subequations}
\begin{figure}
\begin{center}
\includegraphics[scale=0.6]{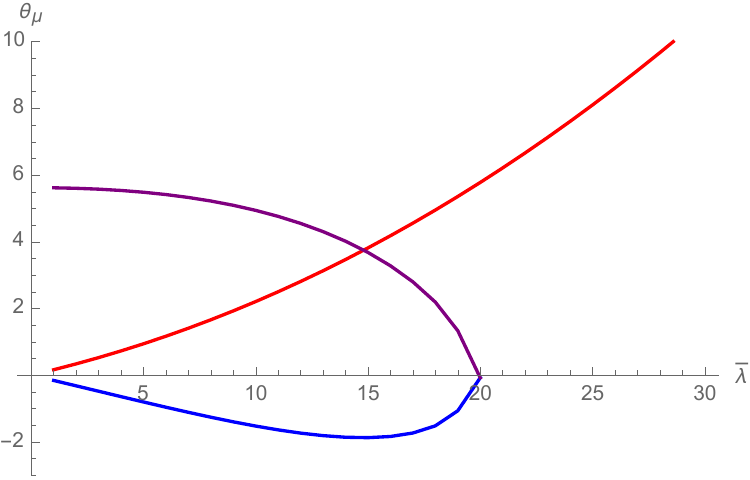}\qquad \quad  \includegraphics[scale=0.6]{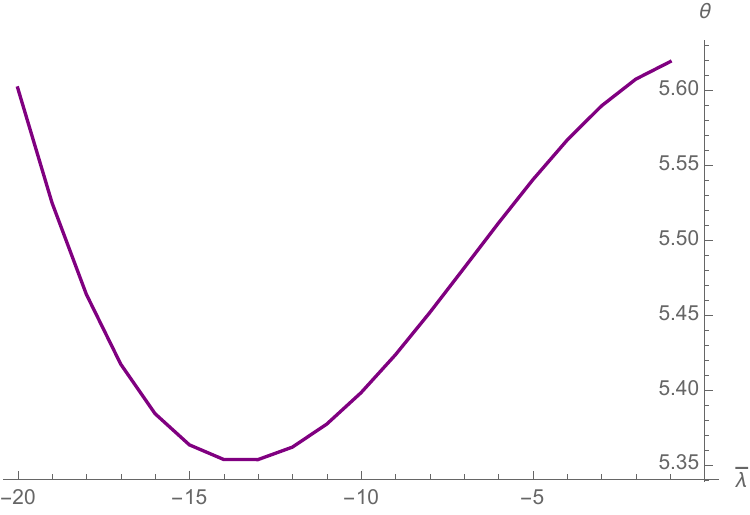}
\end{center}
\caption{Dependency of the critical exponents $\theta_\mu$ ($\mu=1,2,3)$ with respect to $\bar{\lambda}$ along the two pieces of the bounded fixed line (blue and purple curves) and along the unbounded line (red curve). ($\alpha=2/3$).}\label{Expolines}
\end{figure}
\end{subequations}
\end{widetext}

\paragraph{Conclusion.} To summarize this letter, we showed that the effective kinetics occurring in the large $N$ limit for $p$-spin glasses involving a matrix-like disorder allows an intriguing connection with Pisarski's mechanism for sextic theories. For $2$-spin model with a sextic confining potential, we show the existence of a fixed line in the IR, in the domain $\alpha\in [\alpha_c(N),2/3]$. The same kind of phenomenon occurs for the $(2+3)$-spin model, and the transition between the symmetric or broken replica phase is recovered. Nonperturbative techniques seem however to be required to investigate the deep IR regime $\alpha \in (2/3,1)$, which will be the topic of a forthcoming work. The same kind of phenomena is expected for the effective field theories we recently considered for signal detection issues in nearly continuous spectra \cite{lahoche3,lahoche4}, and we plan to investigate this point for future work focusing on the effective potential formalism. Note to conclude that we expect to be protected from the famous BMB instability in this context because the spectra are bounded and the model has no UV singularities.

\end{document}